\documentstyle[12pt]{article}

\catcode`\@=11
\newcommand{\ccite}[1]
{\@ifundefined{b@#1}{\bf ?}{\@nameuse{b@#1}}}

\begin{document}
\setcounter{page}{0}
\baselineskip=20pt

\title{Generalized partition functions, interpolating statistics
and higher virial coefficients}

\author{P. F. Borges\thanks{E-mail address: pborges@if.ufrj.br},
H. Boschi-Filho\thanks{E-mail address: boschi@if.ufrj.br}
and C. Farina\thanks{E-mail address: farina@if.ufrj.br}
\\ 
\\
\small \it
Instituto de F\'\i sica, Universidade Federal do Rio de
Janeiro \\ 
\small \it  Caixa Postal 68528, 
Rio de Janeiro 21945-970, RJ, BRAZIL}

\maketitle
\thispagestyle{empty}

\begin{abstract} 
Starting from determinants at finite temperature obeying an intermediate
boundary condition between the periodic (bosonic) and antiperiodic
(fermionic) cases, we find results which can be mapped onto the ones
obtained from anyons for the second virial coefficient. Using this
approach, we calculate the corresponding higher virial coefficients and
compare them with the results known in the literature.
\end{abstract}

\vskip 3cm

\bigskip
\bigskip

\hfill

\pagebreak

In two dimensional space one finds fractional statistics which depends
on a continuous parameter defined on an interval where bosons correspond
to one end and fermions to the other [\ccite{lmy77}--\ccite{wil90}]. One
of the most striking properties of fractional statistics is the
prediction of fractional charge, {\sl i. e.}, noninteger multiples of
the fundamental electron charge $e$, at low energies inside condensed
matter differently from quarks which can only be probed at very high
energies and are confined to hadrons. In particular, Laughlin explained
the essentials of the fractional quantum Hall effect (FQHE)
\cite{lau83,pgi87} assuming the existence of currents carrying fractional
charges $e/3$ inside two dimensional samples, which has been confirmed
in very recent experiments \cite{pic97,sgj97}.

Different formulations of fractional statistics were proposed, among
which the most fundamental ones are the anyon formulation based on a
quantum mechanical model consisting of charged flux tubes confined to
move on a plane \cite{wil82} and the axiomatic Haldane's fractional
exclusion statistics which is based on very general assumptions on
quantum statistical mechanics in arbitrary dimensions \cite{hal91}.
However, these two formulations are not equivalent in general, as it was
soon noted by Haldane. One way to see this is to make explicitly
calculations, {\sl e. g.}, for the virial coefficients on both
formulations and then compare the results. In the case of the exclusion
statistics it was shown that for a gas in a low-density expansion only
the second virial coefficient depends on the corresponding statistical
parameter while the others are exactly equal to the bosonic and
fermionic values \cite{msh94,iam96}. On the other side, in the charged
flux tube or anyon formulation, the second virial coefficient was found
exactly \cite{asw85} but the higher virial coefficients are not exactly
tractable, apart form the fact that the third virial coefficient has an
exact mirror symmetry in its dependence on the statistical parameter in
respect to its semi-interval \cite{bbk91,sen92}. The calculation of
higher virial coefficients for anyons have been attacked with the aid of
perturbation theory and numerical methods showing that in general, there
are corrections with respect to the corresponding bosonic or fermionic
values [\ccite{mou91}--\ccite{kmm98}].

Recently, we introduced a generalized partition function interpolating
that ones for free relativistic charged bosons and fermions which we
showed to lead, in the nonrelativistic limit, to results which are
mapped onto the conventional ones for anyons, in particular for the
second virial coefficient \cite{bbf98}. In this letter we are going to
use this approach to calculate the corresponding higher virial
coefficients. We find that they are all dependent on the interpolating
parameters, so they only coincide with the results from the exclusion
statistics for the bosonic and fermionic cases. However, these
coefficients are remarkably close to the ones obtained from anyon
calculations.

The route that we are going to take here follows Ref. \cite{bbf98} so
that we use few but essential information which are necessary for a
thermodynamical description of a system whose properties interpolate
those corresponding to fermionic and bosonic gases. These informations
are: first the (relativistic) particles should obey energy conservation
and so despite they are described by some involved specific (say, 2+1
dimensional) equation they should also obey the Klein-Gordon equation.
Hence, the corresponding partition function can be written in terms of the
determinant of this operator; second, this determinant should be
calculated using some appropriate boundary condition; and finally the
partition function is defined by a certain power of this determinant. We
choose a boundary condition that continously interpolates the bosonic
and fermionic cases. Such a condition, in the Euclidean time
compactified to the interval $[0,\beta]$, $\beta$ being the inverse of
temperature, can be written as
\begin{equation}\label{condition} 
\psi(0;\vec x)=e^{i\theta}\psi(\beta ;\vec x), 
\end{equation}

\noindent where $\theta$ is a real continuous parameter. The spatial
analog of this condition was discussed before in the study of the vacuum
structure in gauge theories \cite{rsw78} and to describe the spatial
behavior of anyons in 2+1 dimensions, which acquire an arbitrary phase
when turned around each other \cite{lmy77,wil82,jac90}. In a different
context, a similar condition had been used to interpolate continuously
between the Euclidean and Minkowski metrics to discuss a possible
dynamical origin for Lorentzian spacetime \cite{gre93,cgr94}. Note that
when $\theta=0$ the above condition reduces to a periodic (bosonic)
boundary condition and if $\theta=\pi$ it becomes an antiperiodic
(fermionic) one.

So, we can write a partition
function according to these prescriptions as ($\hbar=c=1$)
\begin{equation}\label{det}
{\cal Z}={\det}^\sigma{(-D^2+M^2)}_\theta, 
\end{equation}

\noindent where $-D^2+M^2$ is the Klein-Gordon operator and the
covariant derivative is defined as $D_\nu=(\partial_0 +
i\mu,\partial_i)$, with a shift in its time component due to a nonzero
chemical potential $\mu$ introduced to describe a finite dinsity of 
(noninteracting) charged
particles of mass $M$ in $(N+1)$ space-time dimensions. The subscript
$\theta$ means that the eigenfunctions of the Klein-Gordon operator
satisfy the boundary condition (\ref{condition}). The power $\sigma$ of
the determinant enables us to correctly describe the partition functions
for charged bosonic ($\theta=0$, $\sigma=-1$) and fermionic
($\theta=\pi$, $\sigma=+1$) gases. 

The determinant (\ref{det}) can be exactly calculated, so that the
corresponding free energy, $\Omega(\beta,\mu)= -(1/\beta)\ln{\cal Z}$,
can be written as \cite{bbf98}:
\begin{equation} 
\Omega(\beta,\mu) = 2 V \sigma 
\left({ M\over 2\pi\beta}\right)^{N+1\over 2} 
\sum_{n=1}^{+\infty}\cos(n\theta)
\left(e^{n\beta\mu}+e^{-n\beta\mu}\right) 
\left({1 \over n}\right)^{N+1\over 2} K_{N+1\over 2} (n\beta M)\;,
\label{Omega}
\end{equation}

\noindent 
where $K_\nu (x)$ is the modified Bessel function of the second kind.
If we particularize the parameters $\sigma$ and $\theta$ to the
bosonic and fermionic cases we shall find precisely the results known
in the literature [\ccite{hwe81}--\ccite{kap89}].

Let us now restrict our analysis to the particular case of
most interest which is the two dimensional ($N=2$) case in the
nonrelativistic limit since this is the context where anyons and their
applications appear. The nonrelativistic limit in this
situation is taken by imposing the low temperature limit $\beta M>>1$,
since we are restricting the allowed energies to be much less than the
rest energy of the particles. Further, we consider only the
particle content of the above free energy, corresponding to the
fugacity $z=\exp{(\beta\mu)}$. So, disregarding the antiparticle 
content with $z^\prime=\exp{(-\beta\mu)}$, we find that
\begin{equation}
\Omega(\beta,\mu)= - {V\over\beta}
\sum_{n=1}^\infty b_n(\beta) z^n,
\end{equation}
where $b_n(\beta)$ are the cluster coefficients:
\begin{equation} \label{b_n}
b_{n}(\beta) = - \; {\sigma M\over 2\pi \beta n^2}\;
\cos(n\theta)\; \exp ( - n\beta M )\;; 
\hskip 1.5cm (\beta M>>1).
\end{equation}

\noindent
Now, using the standard formulae which relate the virial coefficients 
$a_n$ with the cluster ones \cite{hua63,dvo92b} $(a_1 = 1)$,
\begin{equation}\label{ca2}
a_2 = -\; {b_2 \over (b_1)^2}\;; 
\end{equation}
\begin{equation}\label{ca3}
a_3 = -\; 2 {b_3 \over (b_1)^3}\; + 4 {(b_2)^2\over (b_1)^4}\;; 
\end{equation}
\begin{equation}\label{ca4}
a_4 = -\; 3 {b_4 \over (b_1)^4}\; + 18 {b_2 b_3\over (b_1)^5} 
- 20 {(b_2)^3\over(b_1)^6};
\end{equation}

\begin{equation}\label{ca5}
a_5 = -\; 4{b_5\over (b_1)^5}\;+{32b_2b_4+18(b_3)^3\over (b_1)^6}\;
-144 {(b_2)^2 b_3\over (b_1)^7} \;
+ 112 {(b_2)^4\over(b_1)^8}\;;
\end{equation}
\begin{eqnarray}
a_6 = -\; 5{b_6\over (b_1)^6}\;
&+& {60b_3b_4+50b_2b_5\over (b_1)^7}\;
-{315b_2 (b_3)^2+280(b_2)^2b_4\over (b_1)^8} \;
\nonumber\\
\label{ca6}
&+& 1120 {(b_2)^3b_3\over(b_1)^9}\;-672{(b_2)^5\over (b_1)^{10}}
\end{eqnarray}

\noindent and substituting (\ref{b_n}) into these relations, 
we find the virial coefficients corresponding to the generalized
partition function (\ref{det}):
\begin{equation}\label{a2}
a_2 = {\lambda^2 \over 4\sigma}\; (1-\tan^2\theta)\;; 
\end{equation}
\begin{equation}\label{a3}
a_3 = {\lambda^4\over 36\sigma^2}\; 
 (1 + 6\tan^2\theta + 9 \tan^4 \theta)\;; 
\end{equation}
\begin{equation}\label{a4}
a_4 = - {\lambda^6\over 16\sigma^3} \; 
(\tan^2\theta +6 \tan^4 \theta +5\tan^6\theta)\;;
\end{equation}
\begin{equation}\label{a5}
a_5 = - {\lambda^8\over 3600\sigma^4} \; 
(1 -60 \tan^2\theta -1170 \tan^4 \theta 
- 2700 \tan^6\theta -1575 \tan^8\theta)\;;
\end{equation}
\begin{equation}\label{a6}
a_6 = - {\lambda^{10}\over\sigma^5} \; 
({1\over 288} \tan^2\theta +{5\over 24} \tan^4 \theta 
+{145\over 144} \tan^6\theta +{35\over 24} \tan^8\theta
+{21\over 32}\tan^{10}\theta)\;,
\end{equation}

\noindent where $\lambda=\sqrt{2\pi\beta /M}$ is the thermal wavelength
in natural units. Higher order coefficients can be analogously
calculated, but for the moment the above ones are sufficient. 

Before we compare these coefficients with those corresponding to the
anyonic formulation, let us first note that a relation between the
parameters $\theta$ and $\sigma$ can be found, based on the fact that we
do not expect to see any statistical behavior in the one particle
partition function $Z_1=V b_1$. Once $b_1$ is given by (\ref{b_n}) with
$n=1$, we have to impose that $\sigma\cos\theta=constant$ in order to
satisfy that condition. Then, to reproduce correctly the bosonic and
fermionic results we choose
\begin{equation}
\sigma(\theta)=-\cos^{-1}\theta.
\end{equation}

\noindent Using this relation we can
rewrite the virial coefficients (\ref{a2})-(\ref{a6}) for instance, 
as functions of $\sigma=\sigma(\theta)$:
\begin{equation}\label{a2s}
a_2 = {\lambda^2 \over 4}\; ({2\over \sigma} - \sigma)\;; 
\end{equation}
\begin{equation}\label{a3s}
a_3 = {\lambda^4 \over 36}\; ({4\over \sigma^2} - 12 + 9\sigma^2)\;; 
\end{equation}
\begin{equation}\label{a4s}
a_4 = {\lambda^6 \over 16}\; (-{4\over\sigma} +9\sigma -5\sigma^3 )\;;
\end{equation}
\begin{equation}\label{a5s}
a_5 = \lambda^8 \; (-{1\over 225\sigma^4} -{2\over 15\sigma^2} 
+{7\over 10} -\sigma^2 +{7\over 16}\sigma^4)\;;
\end{equation}
\begin{equation}\label{a6s}
a_6 = \lambda^{10} \; (- {1\over 18\sigma^3} +{5\over 8\sigma} 
-{125\over 72}\sigma +{175\over 96}\sigma^3 -{21\over 32}\sigma^5)\;. 
\end{equation}

\noindent So, from eqs. (\ref{a2})-(\ref{a6}) or eqs. 
(\ref{a2s})-(\ref{a6s}) we clearly see that these coefficients are
singular for $\theta=(2k+1)\pi/2$ with $k$ integer or equivalently when
$\sigma=0$. This is not the case of the anyon higher virial coefficients
which have been shown using perturbation and numerical methods to be
finite to all orders except for the second where cusps are present at the
bosonic points [\ccite{sen92}--\ccite{kmm98}]. However, our coefficients
and the usual anyon ones share other interesting properties. First of all,
they naturally reproduce the usual bosonic and fermionic virial
coefficients when we put $\theta=0$ and $\pi$, respectively, so that
\begin{equation}
{a_2}^F_B=\pm{1\over 4}\lambda^2;\qquad
a_3={1\over 36}\lambda^4;\qquad
a_4=0;\qquad
a_5=-{1\over 3600}\lambda^8;\qquad
a_6=0;\;\dots
\end{equation}

\noindent and all coefficients (\ref{a2})-(\ref{a6}) or equivalently
(\ref{a2s})-(\ref{a6s}) are periodic functions of $\theta$, as expected. 
Further, when $\theta$ is away from the singularities, {\sl i. e.}, near
the bosonic $\theta=2k\pi$ and fermionic $\theta=(2k+1)\pi$ points these
expressions give good approximations to the higher virial coefficients of
anyons in a very simple way, as we are going to show bellow. Near these
points we can use the approximation
\begin{equation}
\tan^2\theta\approx\sin^2\theta
\end{equation}

\noindent so that we have the approximate coefficients
\begin{equation}\label{a2a}
a_2 \simeq \pm {\lambda^2 \over 4}\; (1-\sin^2\theta)\;; 
\end{equation}
\begin{equation}\label{a3a}
a_3 \simeq {\lambda^4\over 36}\; 
 (1 + 6\sin^2\theta + 9 \sin^4 \theta)\;; 
\end{equation}
\begin{equation}\label{a4a}
a_4 \simeq \mp {\lambda^6\over 16} \; 
(\sin^2\theta +6 \sin^4 \theta +5\sin^6\theta)\;;
\end{equation}
\begin{equation}\label{a5a}
a_5 \simeq - {\lambda^8\over 3600} \; 
(1 -60 \sin^2\theta -1170 \sin^4 \theta 
- 2700 \sin^6\theta -1575 \sin^8\theta)\;;
\end{equation}
\begin{equation}\label{a6a}
a_6 \simeq \mp {\lambda^{10}} \; 
( {1\over 288} \sin^2\theta +{5\over 24} \sin^4 \theta 
+{145\over 144} \sin^6\theta +{35\over 24} \sin^8\theta
+{21\over 32}\sin^{10}\theta)\;,
\end{equation}

\noindent where the upper signs correspond to fermions and lower to
bosons. Note that the odd coefficients have the same global sign for
bosons and fermions according to Eqs. (\ref{a2})-(\ref{a6}). 

To stablish the relations between these coefficients and those that come
from anyons, let us first write the exact result of Arovas, Schrieffer,
Wilczek and Zee \cite{asw85} for the anyon second virial coefficient: 
\begin{equation}\label{A2}
A_2={\lambda^2\over 4}[\pm 1+(-2\pm 2)|\alpha|-2\alpha^2],
\end{equation}

\noindent where upper signs correspond to fermion, lower to boson based
anyons and $\alpha$ is the statistical parameter, which is assumed to be
a periodic function of another parameter, say an angle $\theta^\prime$. 

Higher anyon virial coefficients have been investigated by several
authors [\ccite{sen92}--\ccite{kmm98}] with different approaches. In
particular, Dasni\`eres de Veigy and Ouvry \cite{dvo92a,dvo92b} using a
highly involved perturbation theory were able to find corrections of
order $\alpha^2$ for the anyon coefficients up to the sixth one:
\begin{equation}\label{A3}
A_3=\lambda^4\left[{1\over 36}+{\alpha^2\over 12}\right];
\end{equation}
\begin{eqnarray}
A_4&=&{\lambda^6\alpha^2\over 16}
\left[{1\over\sqrt{3}}\ln{\sqrt{3}+1\over\sqrt{3}-1}\mp 1\right]
\nonumber\\ \label{A4}
&\simeq& {\lambda^6\alpha^2\over 16}
\left({3\over 4}\mp 1\right);
\end{eqnarray}
\begin{eqnarray}
A_5&=&\lambda^8\left\{-{1\over 3600} 
+ {\alpha^2} \left[{1\over 36}
\mp\left({1\over 6\sqrt{6}}\ln{3+\sqrt{6}\over3-\sqrt{6}}
-{1\over 6\sqrt{3}}\ln{\sqrt{3}+1\over\sqrt{3}-1}\right)\right]\right\}
\nonumber \\ \label{A5}
&\simeq& \lambda^8\left[-{1\over 3600} 
+ {\alpha^2} \left( {1\over 36} \mp {1\over 34}  \right)\right];
\end{eqnarray}
\begin{eqnarray}
A_6&=&{\lambda^{10}\alpha^2}\left[\mp{5\over 576} 
+{5\over 864\sqrt{2}}\ln{\sqrt{2}+1\over\sqrt{2}-1}
+{65\over 288\sqrt{3}}\ln{\sqrt{3}+1\over\sqrt{3}-1}
+{8\over 27\sqrt{5}}\ln{\sqrt{5}+1\over\sqrt{5}-1}
\right.\nonumber\\ 
&&\hskip 1.3cm\left.
-{25\over 48\sqrt{6}}\ln{3+\sqrt{6}\over 3-\sqrt{6}}
+{9\over 32\sqrt{10}}\ln{4+\sqrt{10}\over 4-\sqrt{10}}
\right]
\nonumber\\ \label{A6}
&\simeq&{\lambda^{10}\alpha^2}
\left(\mp{5\over 576} +{1\over 102}\right)
\end{eqnarray}

\noindent following the same notation used in Eq. (\ref{A2}). 
Using very precise Monte Carlo calculations 
these results were later corrected to order $\alpha^4$ 
for the third virial coefficient by Maskevich, Myrheim
and Olaussem \cite{mmo96} 
\begin{equation}\label{A3t}
A_3=\lambda^4\left[{1\over 36}+{\sin^2\theta^\prime\over 12 \pi^2}
-{\sin^4\theta^\prime\over (621\pm 5)\pi^4}\right];
\end{equation}

\noindent where $\sin\theta^\prime=\pi|\alpha|$, 
and more recently by Kristofersen {\sl et.
al.} to the fourth anyon virial coefficient \cite{kmm98}:
\begin{equation}\label{A4t}
A_4=\lambda^6\left[{\sin^2\theta^\prime\over 16\pi^2}
\left({1\over\sqrt{3}}\ln{(\sqrt{3}+2)}+\cos\theta^\prime\right)
+\sin^4\theta^\prime\left(c_4+d_4\cos\theta^\prime\right)\right],
\end{equation}

\noindent  where 
$c_4 = -0.0053\pm 0.0003$ and $d_4 = -0.0048\pm 0.0009$.

Note that there is a small difference between the expressions of $A_4$,
Eqs. (\ref{A4}) and (\ref{A4t}), in respect to the coefficient of term
$\sin^2\theta^\prime$ because of the substitution of $\mp 1$ by
$\cos\theta^\prime$. This does not modify our analysis since we are
considering $\theta^\prime$ near $\pi$ or zero. More generally, it seems
to be a basic difference between these two approaches to higher anyon
virial coefficients. In numerical simulations one usually works with
single smooth functions which do not distinguish the statistics, while
in the perturbation theory, as expressed in coefficents $A_4$ and $A_6$,
Eqs. (\ref{A4}) and (\ref{A6}), these coefficients are statistics
dependent as it happens to the well known Arovas {\sl et. al.} result, Eq.
(\ref{A2}).

So, near the fermionic point in our formulation ($\theta\approx\pi$) and
imposing the equality between (\ref{a2a}) and (\ref{A2}) with fermion
signs, we obtain
\begin{equation}\label{relf}
\sin^2 \theta\simeq 2\alpha^2
\simeq {2\over \pi^2}\sin^2\theta^\prime
\end{equation}

\noindent which shows that our virial coefficients , Eqs.
(\ref{a3a})-(\ref{a6a}), are very close to the ones coming from anyons, 
Eqs. (\ref{A3})-(\ref{A4t}). 
An analogous situation occurs if we start near the bosonic point
($\theta\approx 0$) and we identify
\begin{equation}\label{relb}
\sin^2\theta\simeq 4|\alpha|-2\alpha^2
\simeq {4\over\pi}|\sin\theta^\prime|
-{2\over\pi^2}\sin^2\theta^\prime
\end{equation}

\noindent 
so that similar results are obtained but since the above relation
include linear terms in $|\sin\theta^\prime|$ one would also find odd
powers of this term in the analysis which did not occur for the
coefficients near the fermionic point.

These results are in good qualitative agreement with the perturbative
and numerical ones for anyons but quantitatively
the numerical factors in our coefficients are greater than those
appearing in Refs. \cite{dvo92a,dvo92b,mmo96,kmm98}. 
However, it is important to emphasize that the our 
coefficients have the expected properties like the exact mirror
symmetry \cite{sen92} and also contain the particular fermionic and
bosonic cases when we choose appropriately the parameter $\theta$.  

Another good test for our coefficients is to compare them with the ones
coming from the average field formulation
[\ccite{lau88}--\ccite{cww89},\ccite{wil90}] where anyons are supposed
to be a large collection of noninteracting bosons
($\nu=\theta^\prime/\pi$) or fermions ($\nu=\theta^\prime/\pi-1$)  
of density $\rho$ in an
average magnetic field, $\bar{B}=\nu\rho\Phi_0$, where $\Phi_0=h/e$ is
the elementary quantum flux. In this case, the
virial expansion reads
\begin{equation}
\beta P=\rho\pm{1\over 4}\lambda^2\rho^2+{1+3\nu^2\over
36}\lambda^4\rho^3\mp{\nu^2\over 16}\lambda^6\rho^4 
-{1-100\nu^2+5\nu^4\over 3600}\lambda^8\rho^5+\cdots.
\end{equation}
Our results (\ref{a2a})-(\ref{a6a}) are remarkably close to these ones, 
suggesting that our results may be an intermediary formulation between
the conventional ones for anyons and the average field approach.

We can go even further in the calculation of higher virial
coefficients within our approach since we already found the
cluster coefficients exactly. For simplicity, we just present the basic
behavior of the n-th virial coefficient:
\begin{equation}
a_n = (1-n) {b_n \over (b_1)^n} + \sum_i c^n_i \prod_j^{n-1} 
(a_j)^{d^n_i}\;, 
\end{equation}

\noindent where $c^n_i$ and $d^n_i$ are well known constants that can be
determined from standard Eqs. (\ref{ca2})-(\ref{ca6}) and so on. 
Using eq. (\ref{b_n}) one finds
\begin{equation}
a_n = \left(-{\cos\theta\lambda^2\over 2n}\right)^{n-1} \; 
 \sum_{i=1}^n k_i^n \tan^{2i-2} \theta\;, 
\end{equation}

\noindent where the constants $k^n_i$ can be determined from Eqs.
(\ref{ca2})-(\ref{ca6})..., combined with (\ref{b_n}). In particular,
$(-1/2n)^{n-1}k^n_1$ are identified with the Bernoulli numbers. Note
that these polynomials in $\tan^2\theta$ with even $n>2$ vanish for
$\theta=0$ and $\pi$, reproducing correctly the usual bosonic and
fermionic virial coefficients in two dimensions.

It is the hope of the authors that the higher virial coefficients
presented here, which interpolate the bosonic and fermionic cases, can
perhaps shed some light on the thermodynamics of fractional statistics. It
should be remarked that we discussed here the free case only. However, in
more realistic models one should take into account the coupling between
the particles and/or the interaction with external fields. Another point
of interest is the possibility of connection of these results with a more
fundamental theory starting from a Lagrangian density with bosonic or
fermionic fields coupled to a Chern- Simons term at finite temperature to
see if the quasiperiodic conditions would come directly from the gauge
symmetry of the problem.  The discussion of these points are presently
under investigation and will be reported elsewhere. Let us also mention
that our approach is independent of the space-time dimension, as it
happens in Haldane's exclusion statistics, although here we discussed the
particular case of $N=2$ in more detail to stress its relations with the
anyon formulation on the basis of the virial expansion. Within this
context, it might be interesting to discuss the relation of our approach
with systems in other dimensions, {\sl e. g.}, the one-dimensional cases
presenting exotic statistics. 


\bigskip

H.B.-F. thanks D. Bazeia for reading a preliminary version of the
manuscript and for useful suggestions on it, and the hospitality of the
Center for Theoretical Physics, MIT where part of this work was done.  The
authors H.B.-F. and C.F. were partially supported by CNPq (Brazilian
agency).

\end{document}